\documentclass[a4paper,12pt]{article}
\usepackage[all]{xy}
\usepackage{amsfonts}
\usepackage{amssymb}
\usepackage{epsfig}
\usepackage{graphicx}
\usepackage{citesort}

\usepackage[latin1]{inputenc}

\def\beq{\begin{equation}}
\def\eeq{\end{equation}}
\def\bea{\begin{eqnarray}}
\def\eea{\end{eqnarray}}

\def\u1{\widehat{U(1)}}
\def\su2{\widehat{SU(2)}_1}

\catcode`\@=11
\def\numberbysection{\@addtoreset{equation}{section}
        \def\theequation{\thesection.\arabic{equation}}}

\begin{document}

\begin{titlepage}
\begin{center}
\hfill  \quad q-bio/yymmnnn \\
\vskip .5 in {\Large\bf  Thermodynamics of a model}

{\Large\bf for RNA folding} \vskip 0.3in

Mat\'{\i}as G. dell'Erba\\
{\em Departamento de F\'{\i}sica, Facultad de Ciencias Exactas y
Naturales, Universidad Nacional de Mar del Plata, Funes 3350,
(7600), Mar del Plata, Argentina}
\\
\vskip 0.3in Guillermo~R.~Zemba\footnote{ Member of CONICET,
Argentina.}
\\
{\em
Facultad de Ciencias Fisicomatem\'aticas e Ingenier\'{\i}a,
Universidad Cat\'olica Argentina, Av. A. Moreau de Justo 1500, Buenos Aires, Argentina}
\\
and
\\
{\em Departamento de F{\i}sica, C.N.E.A.
Av.Libertador 8250, (1429) Buenos Aires, Argentina}
\end{center}
\vskip .5 in
\begin{abstract}
\noindent We analyze the thermodynamic properties of a simplified
model for folded RNA molecules recently studied by G. Vernizzi, H.
Orland, A. Zee (in {\it Phys. Rev. Lett.} {\bf 94} (2005) 168103).
The model consists of a chain of one-flavor base molecules with a
flexible backbone and all possible pairing interactions equally
allowed. The spatial pseudoknot structure of the model can be
efficiently studied by introducing a $N \times N$ hermitian random
matrix model at each chain site, and associating Feynman diagrams
of these models to spatial configurations of the molecules. We
obtain an exact expression for the topological expansion of the
partition function of the system. We calculate exact and
asymptotic expressions for the free energy, specific heat,
entanglement and chemical potential and study their behavior as a
function of temperature. Our results are consistent with the
interpretation of $1/N$ as being a measure of the concentration of
$\rm{Mg}^{++}$ in solution.
\end{abstract}



\vfill
\end{titlepage}
\pagenumbering{arabic}


The applications of mathematical and statistical mechanics
techniques to study suitable biological problems has been a
successful area of recent research interest \cite{Gen1,Gen2,Gen3}.
In particular, the study of the spatial and topological
(pseudoknot) structure of DNA and RNA molecules is a successful
example of the above
\cite{Muller,Zee1,Zee2,Zee3,Zee4,Zee5,David,Tinoco}. A RNA
molecule is a heteropolymer strand made up of four types of
nucleotides: uracil ($U$), adenine ($A$), guanine ($G$), and
cytosine ($C$). The sequence of these bases from the $5'$ to the
$3'$ end defines the primary structure of the molecule. In
solution, at room temperature, different bases can pair with each
other by means of saturating hydrogen bonds to give the molecule a
stable shape in three dimensions, with $U$ bonding to $A$, $C$ to
$G$, and wobble pair $G$ to $U$, all with different interaction
energies. This last interaction (non Watson-Crick base pair)
together with triplets, quartets, etc. has a important role in
fold of the RNA molecule \cite{Tertiary,Multi,GU,Zhang}. The
effect of stacking interactions also contributes to the stability
of the molecule, making sets of adjacent bonds twist into the
familiar Watson-Crick helices. Among all possible structures that
arise from interaction between the bases, one defines the
secondary structures of a RNA molecule as all structures which are
represented by planar arc diagrams, that is, no crossing of arcs
in a representation resembling a Feynman diagram. When the
diagrams are non-planar, one says that a RNA molecule contains one
or more pseudoknots (see \cite{knots1,knots2} for a general
definition and \cite{Zee6} for a discussion on
the planarity of the diagrams). Finally, one defines the tertiary
structure of RNA as the actual spatial three-dimensional
arrangement of the base sequence.

Several methods have been successfully used to study the folding
dynamics of RNA molecules in various conditions. Some of these are
based on statistical mechanics models, which usually avoid the
complexities related to the dynamical evolution of the real world
molecules, but allowing for a simple, kinematical treatment of the
proposed models. Therefore, the study of these models can shed
light on the intricate molecular dynamics and is our main
motivation for considering them. In this paper, we study a
simplified model of a RNA-like molecule considered in \cite{Zee1}
in which the geometric degrees of freedom of the system, such as
the stiffness and the sterical constraints of the chain,  are not
taken into account. In addition, they consider that all pairs of
bases interact with a common pairing strength
(the assumption of neglecting disorder along the sequence is
actually a classic approximation \cite{deGennes}).
Moreover, the model keeps the fundamental
property of saturation of the interactions, that is, given a base
in the chain it can interact (following the rules mentioned above)
with only one another base at a time. The study of this model is
interesting in itself and has motivated some interesting work in
the literature, including case of the planar diagram limit
(no pseudo-knots) \cite{model}\cite{model2}\cite{model3}, and
the study of the tertiary structure of
the RNA molecule (see, for instance
\cite{Zee1,Zee2,Zee3,Zee4,Zee5}).
Moreover, the model allows for
some exact calculations including the partition function, the
specific heat, and some other thermodynamical and physical
quantities.Therefore, the study of the physical properties of this
model could be considered as a first approximation or a limit case
for more realistic RNA models. A natural extension of the model in
\cite{Zee1} towards more realistic ones (for example, including
different interactions between the bases) could be developed by
simple modifications of a matrix potential.

\begin{figure}[h]
\includegraphics[trim=0mm 43mm 17mm 10mm,clip,width=12cm]{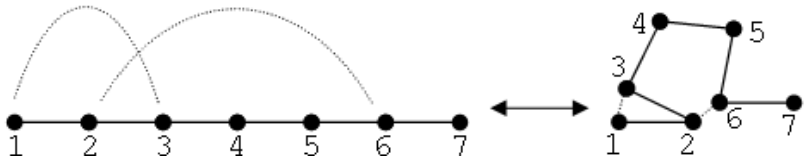}
\caption{Arc diagram representation of the interacting pairs
$(1,3)$ and $(2,6)$ and its respective folded diagram.}
\label{Plegamiento}
\end{figure}

The authors of \cite{Zee1} consider a system of $L$ molecules
(nucleotide bases) forming a lineal macromolecule with the shape
of a chain. They do not describe the formation of the backbone,
but only the interaction between links of the chain that produce
the folding of the RNA macromolecule (see Fig.\ref{Plegamiento}).
Each base can interact through an attractive force with any other
base of the chain. However, once a given molecule has paired with
other, it will not interact again with another. In this case, it
is said that the interaction between these bases saturate.

Although the bases that form the RNA molecule interact with
different pairwise energies, a first simple approximation is to
consider that this energy is the same for all bonds, and that any
paring of bases is assumed to be feasible. This amounts to
considering just one type of base and no further selection rules.
Note that if all energies are equal, then the Boltzmann factors
$v=e^{-\frac{\epsilon}{\kappa T}}$ (where $T$ is the absolute
temperature and $\kappa$ is the Boltzmann constant which we will
equal to one) are equal as well.
 The configurational partition function of a molecule of
size $L$ in the model in \cite{Zee1} can be written as

 \beq
Z(L,N,T)=\frac{1}{N}\langle{\rm
tr}(1+\frac{1}{L^{1/2}}\varphi)^{L}\rangle=\frac{\int d\varphi\
e^{-\frac{N}{2Lv}{\rm tr}\varphi^{2}}\frac{1}{N}{\rm
tr}(1+\frac{1}{L^{1/2}}\varphi)^{L}}{\int d\varphi\
e^{-\frac{N}{2Lv}{\rm tr}\varphi^{2}}}, \label{Zvalormedio}
 \eeq
\noindent
where $\varphi$ is a random $N \times N$ hermitian matrix and $Z$
depend of $T$ through $v$. Note that the simple form of
(\ref{Zvalormedio}) is only a consequence of the symmetry of the
matrix potential that reduces the original integration over $L$
matrices to one integration over $\varphi$ \cite{Zee1}. From the
theory of random matrices (see for example \cite{Mehta}, pag.
140-2) it follows that
 \beq
\langle {\rm tr}\varphi^{2k}
\rangle=\frac{(2k)!}{k!}\left(\frac{Lv}{2N}\right)^{k}
\sum_{j=0}^{k}
\left(%
\begin{array}{c}
  k \\
  j \\
\end{array}%
\right)
\left(%
\begin{array}{c}
  N \\
  j+1 \\
\end{array}%
\right)2^{j} \ ,
 \eeq
\noindent where all averages are performed with respect to the
gaussian measure $d\varphi\ e^{-\frac{N}{2Lv} {\rm
tr}\varphi^{2}}$. Replacing this into (\ref{Zvalormedio}) and
taking into account that $\langle {\rm tr}\varphi^{k} \rangle=0$
for $k$ odd, we arrive at:
 \beq
Z(L,N,T)=\sum_{k=0}^{[L/2]}\sum_{j=0}^{k}\left(%
\begin{array}{c}
  L \\
  2k \\
\end{array}%
\right)
\left(%
\begin{array}{c}
  k \\
  j \\
\end{array}%
\right)
\left(%
\begin{array}{c}
  N \\
  j+1 \\
\end{array}%
\right) \frac{(2k)!}{2^{k-j}\ k!\ N^{k+1}}\ v^{k}
\label{PartFunct1}
 \eeq
\noindent

where the symbol $[L/2]$ means the integer part of $L/2$. From
(\ref{PartFunct1}) we may compute (for each $L$) $Z$ exactly, as a
function of $N$ and $T$. The large-$N$ asymptotic expansion of $Z$
has a well-known topological meaning \cite{'t Hooft}: the power of
$v$ is the number of arcs in the diagram, and the power of
$1/N^{2}$ is the genus $g$ of the diagram. It is therefore
convenient to write 
(\ref{PartFunct1}) in following form:

 \beq
Z(L,N,T)=\sum_{k=0}^{[L/2]} d_{k}(L,N)\ e^{-\epsilon_{k}/T}
 \label{PartFunctTrad}
 \eeq
\noindent
where $\epsilon_{k}= k \epsilon$ and
 \beq
d_{k}(L,N)=\sum_{j=0}^{k} \left(%
\begin{array}{c}
  L \\
  2k \\
\end{array}%
\right) \left(%
\begin{array}{c}
  k \\
  j \\
\end{array}%
\right) \left(%
\begin{array}{c}
  N \\
  j+1 \\
\end{array}%
\right)\frac{(2k)!}{2^{k-j} \ k! \ N^{k+1}}
 \label{Deg}
 \eeq
\noindent

From (\ref{PartFunctTrad}) we see that the spectrum of the system
has $[L/2]+1$ possible energies
$0,\epsilon,2\epsilon,\ldots,[L/2]\epsilon$ and the degeneracy of
the $k-th$ level is $d_{k}(L,1)$. For example, for $L=7$ the
maximum energy of a configuration is $3 \epsilon$ and there are
$d_{3}(7,1)=105$ different configurations with that energy.
Moreover, considering (\ref{Deg}) as a function of $N$ yields its
topological information, {\it e.g.}, for $L=7$, $d_{3}(7,N)=35+70\
1/N^{2}$, which means that out of the total $105$ configurations
with $3 \epsilon$, $35$ have genus $0$ and $70$ have genus $1$.


Next, we calculate the partition function in the large $N$ limit,
which is the planar limit, using the results of \cite{Staudacher}.
For completeness, we quote here the results relevant for ours. We
define the resolvent $W(p)$:
 \beq
W(p)=\frac{1}{N}\left \langle {\rm tr} \left(
\frac{1}{p-\varphi}\right)\right
\rangle=\sum_{n=0}^{\infty}\left(\frac{1}{p}\right)^{n+1}W_{n}\ ,
 \eeq
\noindent
where $p$ is a complex variable, and
\beq
W_{n}=\frac{1}{N}\langle {\rm tr} \varphi^{n} \rangle \ .
 \eeq
\noindent In the large $N$ limit, the resolvent is given by the
solution to the following equation called Pastur's equation
\cite{Pastur} (in the limit $g\rightarrow0$ and $c\rightarrow0$,
in \cite{Staudacher}):
 \beq
W^{2}(p)-p \hspace{0.1cm} W(p)+1=0 \ ,
 \eeq
\noindent which is:
 \beq
W(p)=\sum_{k=0}^{\infty}C_{k}\frac{1}{p^{2k+1}}=\sum_{k=0}^{\infty}\frac{\left(%
\begin{array}{c}
  2k \\
  k \\
\end{array}%
\right)}{k+1}\frac{1}{p^{2k+1}}\ , \label{Catalan}
 \eeq
\noindent where $C_{k}$ are the Catalan numbers \cite{Gradshteyn}.
From (\ref{Catalan}) we obtain
 \beq
\langle {\rm
 tr}\varphi^{2k}\rangle=\frac{\left(%
\begin{array}{c}
  2k \\
  k \\
\end{array}%
\right)}{k+1}L^{k}v^{k} \label{ValorMedioInf}
 \eeq
\noindent With (\ref{ValorMedioInf}) we write down the partition
function in the large $N$ limit. We consider the case $N=1$ for
comparison purposes as well:
 \beq
 Z(L,N\rightarrow\infty,T)=\sum_{k=0}^{[L/2]} \frac{L!}{(L-2k)!\ k!\ (k+1)!}\ v^{k}
 \eeq

 \beq
 Z(L,N=1,T)=\sum_{k=0}^{[L/2]} \frac{L!}{(L-2k)!\ k!\ 2^{k}}\ v^{k}
 \eeq
\noindent Note that both expressions  for $Z$ are very similar,
except for the factors $(k+1)!$ and $2^{k}$ in the denominators of
the expansion coefficients. Noting that the first factor is larger
than the second, we conclude that $Z(N\rightarrow\infty) \leq
Z(N=1)$. The interpretation of this result is clear if we recall
that, for $v=1$, $Z(N\rightarrow\infty)$ counts the planar
diagrams only, whereas $Z(N=1)$ counts both the planar and
non-planar diagrams \cite{Zee1,Zee2,Zee3,Zee4}. Morevover, we
verify that both partitions functions coincide for values of $L$
smaller than $3$ as they should given that all diagrams are planar
in these cases.
 Furthermore, in these two limiting cases, the partition function can be written as
in terms of hypergeometric functions:
 \beq
Z(L,N=1,T)={_2F_{0}}(-\frac{L}{2},-\frac{L}{2}+\frac{1}{2};2v)\ ,
\label{hyper20}
 \eeq
 \beq
Z(L,N\rightarrow\infty,T)={_2F_{1}}(-\frac{L}{2},-\frac{L}{2}+\frac{1}{2};2;4v)\
, \label{hyper21}
 \eeq
\noindent where
${_pF_{q}}(\overrightarrow{a};\overrightarrow{c};z)=\sum_{k=0}^{\infty}\frac{(a_{1})_{k}\ldots(a_{p})_{k}}{(c_{1})_{k}\ldots(c_{q})_{k}}\frac{z^{k}}{k!}$
and $(a)_{k}=\frac{\Gamma(a+k)}{\Gamma(a)}$ are the $k$-order
Pochhammer symbols. We remark here that the results
(\ref{hyper20}) and (\ref{hyper21}) are exact.


As we mentioned before, the power of $1/N^{2}$ yields the genus
$g$ of the diagram, that is, the minimum number of handles of the
surface on which the diagram can be drawn without crossings. From
table of values of $Z$ for the smallest values of $L$ in
\cite{Zee1} we notice, for instance, that for $L=5$ the number of
planar diagrams on the sphere is $21$ and the number of non-planar
diagrams that can be drawn on a torus without crossings is $5$.
Next, we would like to write $Z(L,N,T)$ in the form of a
topological expansion \cite{'t Hooft,Zee1,Zee2}, that is, as a
power series in the variable $1/N^{2}$:
 \beq
 Z(L,N,T)=\sum_{g=0}^{\infty} z_{g}(L,T) \frac{1}{N^{2g}}\ ,
 \label{ExpTop}
 \eeq
\noindent where $z_{g}(L,\infty)$ is, for a molecule of size $L$,
the number of planar diagrams in a topological surface of genus
$g$. For the example of the previous paragraph, we have
$z_{0}(5,\infty)=21$ and $z_{1}(5,\infty)=5$. Note that
$z_{g}(L,T)$, as a function of $T$, is the partition function of
the system living on the topological surface of genus $g$. In
order to bring the partition function to the form (\ref{ExpTop}),
we first define the auxiliary function:

 \beq
 G_{k}(N)=\sum_{j=0}^{k}
 \left(%
\begin{array}{c}
  k \\
  j \\
\end{array}%
\right)
\left(%
\begin{array}{c}
  N \\
  j+1 \\
\end{array}%
\right)\frac{2^{j}}{N^{k+1}}. \label{FunAux}
 \eeq
\noindent
This function contains all the $N$ dependence of
(\ref{PartFunct1}). Below, we write the binomial coefficient as
 \beq
\left(%
\begin{array}{c}
  N \\
  j+1 \\
\end{array}%
\right)=\frac{1}{(j+1)!}N(N-1)\cdots(N-j)=\frac{1}{(j+1)!}\sum_{m=0}^{j+1}S_{j+1}^{(m)}N^{m}\ ,
\label{Stirling}
 \eeq
\noindent 
where $S_{j}^{(m)}$ is the Stirling number of the first
kind \cite{Gradshteyn,Stirling} with parameters $m,j$ (in turn, we
define $S_{j}^{(m)}=0$ if $m>j$ or if $j\leq0$). Replacing
(\ref{Stirling}) in (\ref{FunAux}), we obtain
 \beq
 G_{k}(N)=\sum_{j=0}^{k}\left(%
\begin{array}{c}
  k \\
  j \\
\end{array}%
\right)\frac{2^{j}}{(j+1)!}
\sum_{m=0}^{j+1}\frac{S_{j+1}^{(m)}}{N^{k+1-m}},
 \eeq
Now, if we want to obtain the $O(1/N^{2g})$ of $G_{k}(N)$ (we
indicate this by $G_{k}^{(2g)}(N)$), we must require that
$k+1-m=2g$, then $j=k-2g,k-2g+1,\ldots,k$. To obtain all orders of
$N$ we must add all possible values of $g$:
 \beq
G_{k}(N)=\sum_{g=0}^{\infty}G_{k}^{(2g)}(N)=
 \sum_{g=0}^{\infty}\sum_{j=k-2g}^{k}\left(%
\begin{array}{c}
  k \\
  j \\
\end{array}%
\right)\frac{2^{j}}{(j+1)!}\frac{S_{j+1}^{(k+1-2g)}}{N^{2g}},
 \eeq
replacing in (\ref{PartFunct1}) we obtain (\ref{ExpTop}) with
$z_{g}(L,T)$ given by: \beq
z_{g}(L,T)=\sum_{k=0}^{[L/2]}\sum_{j=k-2g}^{k}
 \frac{L! \ 2^{j-k} \ S_{j+1}^{(k+1-2g)}}{(L-2k)! \ (k-j)! \ j! \
 (j+1)!} \
e^{- \epsilon_{k}/T}. \label{ZxTop} \eeq 
\noindent
In the limit $T\rightarrow\infty$, $z_{g}(L,T)$ coincides with $a_{L,g}$ of
\cite{Zee1}. Using the above mentioned property of the Stirling numbers,
we see from (\ref{ZxTop}) that the maximum
genus of a diagram for a given $L$ is $[L/4]$, therefore
$g\leq[L/4]$. An analysis of the $T$-dependent phase transition from
the topological expansion of the partition function will be given
in a future publication \cite{dellerba}.



In the rest of this letter, we analyze the thermodynamic
properties implied by the partition functions (\ref{PartFunct1})
by studying some interesting observables. We start by considering
the 'entanglement' (non-local two-point correlation function)
between two bases of the the chain in our model. For that, we use
the following definition for the correlation between the molecules
$i$ and $j$ ($i<j$) of the chain of size $L$

\beq
 \langle i,j \rangle=\frac{Z(\overline{j-i},N,T)}{Z(L,N,T)}, \label{corr-def}
\eeq \noindent
where $Z(\overline{j-i},N,T)$ is the partition function for
the molecule including bases $i$ up to  $j$. For the case of
periodic boundary conditions we have
$Z(\overline{j-i},N,T)=Z(j-i+1,N,T)$. In the low temperature limit
the partition function becomes independent of $N$ (in this limit
only configurations up to two bases interacting are possible, it
is, planar configurations). Therefore yields the same result for
both the $N=1$ and $N\rightarrow\infty$ cases:

 \beq
\langle i,j
\rangle=\frac{1+\frac{1}{2}x(x+1)v}{1+\frac{1}{2}L(L-1)v}\simeq
1-\frac{1}{2}(L^{2}-x^{2})e^{-\beta \epsilon}\ ,
\label{corr-vCERO}
 \eeq
\noindent
with $L>>1$, $x=j-i>>1$ (provided $\beta
\epsilon>>\ln(L^{2}/12)+\ln(2+1/N^{2})$). The physical behavior of
this observable can be obtained by considering the situation where
$x$ is large, yielding $\langle i,j \rangle \simeq x^2 $, which
can be interpreted as a signal of confinement (long-range order
with critical exponent $-2$). This behavior is coherent with the
interpretation of the model as describing a folded RNA molecule.
Note that the exponent $(-2)$ for the long range order coincides
with the value found in \cite{model3,expo}.


Going ahead with our study of observables of the model, we now
calculate the normalized free energy $f$, or free energy per
molecule, in the limit of low temperature:
 \beq
f(L,N,T)=\frac{F(L,N,T)}{L}=-\frac{1}{\beta L}\ln(Z(L,N,T))
 \eeq
\noindent where $F$ is the free energy of the system. For  $T<<1$,
both the $N=1$ and $N\rightarrow\infty$ cases yield:
 \beq
f(\beta>>1)=-\frac{1}{2}(L-1)\frac{e^{-\beta \epsilon}}{\beta}
 \eeq
\noindent
Furthermore, for the special case of $v=\frac{1}{4}$ (for which
$\beta=\ln 4/\epsilon$) we obtain
 \beq
f(N\rightarrow\infty,v=1/4)=-\frac{1}{\beta}\ln(2)
 \eeq

\begin{figure}[h]
\epsfysize=7cm \centerline{\epsfbox{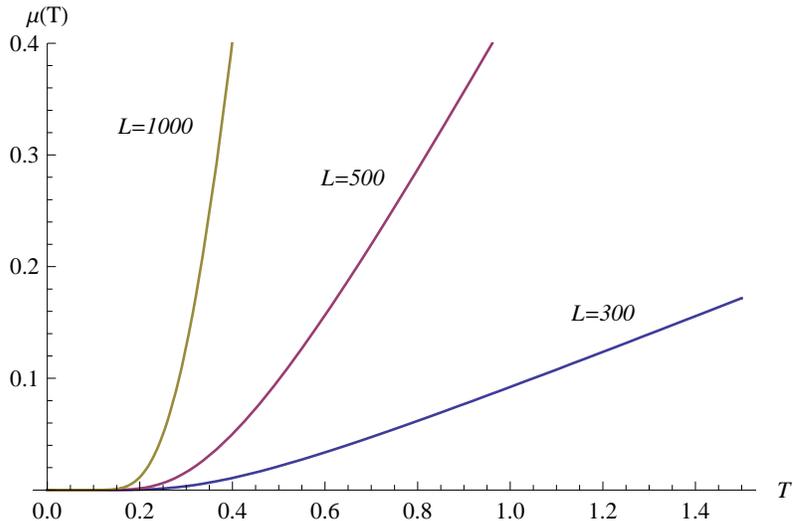}} \caption{Chemical
potential as a function of the temperature for $N=100$.}
\label{mu}
\end{figure}

Next, we define the chemical potential of the model as:

 \beq
 \mu(L,N,T) =\frac{\partial F(L,N,T)}{\partial N} =
-\frac{T}{Z(L,N,T)}\frac{\partial Z(L,N,T)}{\partial N}
 \eeq
\noindent
The interpretation of $\mu$ is the following: we consider that
there is a gas of $N$ 'particles' in the internal space of the
random matrices at each site of the chain of size $L$. The
chemical potential measures the response of the system to a change
in the size $N$ of the matrix $\varphi$. On the other hand, the
concentrations of secondary and tertiary structures can be
separated experimentally by varying the concentration of
$\rm{Mg^{++}}$ ions in solution \cite{Mg++,Zee2,Zee5}; in the
original model \cite{Zee1}, one can assign this role of regulation
to $N$, as it is mentioned in \cite{Zee5} and can be seen from
(\ref{ExpTop}) that this dependency is how $1/N \sim
[\rm{Mg^{++}}]$. Therefore, the chemical potential $\mu$ can be
considered as a measure of the influence of the concentration of
$\rm{Mg^{++}}$ ions in solution on the system. From (\ref{ExpTop})
we see that, in the large $N$ limit, $Z$ is $O(1)$ and $\partial
Z/\partial N$ is $O(1/N^{3})$, therefore $\mu$ is $O(1/N^{3})$ in
the form:
 \beq
\mu (L,N,T)= \frac{2 T}{N^{3}} \frac{z_{1}(L,T)}{z_{0}(L,T)} +
O(1/N^{5})
 \eeq
\noindent
In the large $N$ limit, we obtain 
the partition functions on the sphere and on the torus,  $z_{0}$ and $z_{1}$
respectively, and write down the chemical potential  
(see \cite{Stirling} for explicit expressions of $S_{n}^{(n-\alpha)}$) 
:
\beq
\mu (L,N,T) \simeq \frac{T}{6 N^{3}} \langle k^{3}-k \rangle_{0},
 \eeq
\noindent
where the averages are defined as:
\beq
\langle r_{k} \rangle_{0}
(L)=\left.\sum_{k=0}^{[L/2]}\frac{e^{-\epsilon_{k}/T}
r_{k}}{(L-2k)! \hspace{0.1cm} k! \hspace{0.1cm} (k+1)!
\hspace{0.1cm}} \right/
\sum_{k=0}^{[L/2]}\frac{e^{-\epsilon_{k}/T}}{(L-2k)!
\hspace{0.1cm} k! \hspace{0.1cm} (k+1)!},
 \eeq
\noindent
and are labelled by the subindex $0$ in order to distinguish
them from the previously defined ones. Using numerical
calculations, it can be seen that for $T >> 1$, $\langle k^{3}\rangle_{0}>>
\langle k \rangle_{0}$ and $\langle k^{3}\rangle_{0}$ is
independent of $T$, then:
 \beq
\mu (L,N,T>>1) \simeq \frac{\langle k^{3}\rangle_{0}}{6 N^{3}}
\hspace{0.05cm} T, \label{muT}
 \eeq
\noindent
whereas for $T << 1$, we see that the dependence of  $\mu$ on
$T$ is given by:
 \beq
\mu (L,N,T<<1)\simeq \frac{(L-3)_4}{12 N^{3}} \hspace{0.05cm} T
e^{-2 \epsilon /T}. \label{mut}
 \eeq
\noindent
The limits we have just discussed are summarized graphically 
in fig. \ref{mu}. For any value
of the temperature, the system will tend to configurations with
large $N$, because this minimizes the chemical potential. In this
regime of $N$, the concentration of positive ions in solution
is small, and the configurations of the molecules
will tend to be planar.

The previous expressions for $\mu$ lead us to a consistent physical
interpretation of the parameter $N$. We recall that
\cite{Pathria}:

 \beq
\frac{\partial S}{\partial N}= - \frac{\mu}{T} ,
 \eeq
\noindent
 where $S$ is the entropy. From equations (\ref{muT}) and (\ref{mut}),
 one can see that $S \propto 1/N^{2}$ in both $T \rightarrow 0$ and $T \rightarrow
 \infty$ limits. Therefore, the entropy vanishes for $N \rightarrow
 \infty$, which is also the limit in which the topology of the
 molecule is spherical (by the topology of a
 molecule, we mean topology of the Feynman diagrams associated
 with the configuration of the molecule). This
 suggests that $S$ could be considered as an indicator of
 the spatial topological configurations of the
 molecule. One can argue that the genus of the molecule is largely
 determined by the conditions of the surrounding medium, such as
 the concentration of $\rm{Mg^{++}}$ ions. The competition
 between the interaction of a given base molecule with other
 molecules in the chain and with the ions of the medium
 regulates the
 folding of the chain and therefore, its genus. One may
 assume that the concentration of ions in the medium should be
 monotonous functions of $1/N$. Therefore, we arrive to the conclusion
 that the 'internal' parameter $N$, introduced by hand as a convenient
 variable for organizing the topological configurations, could be given
 the physical interpretation of representing the inverse quantity of
 the ion concentration of the medium \cite{Zee2,Zee5,Zee6}.

\begin{figure}[h]
\epsfysize= 5cm \centerline{\epsfbox{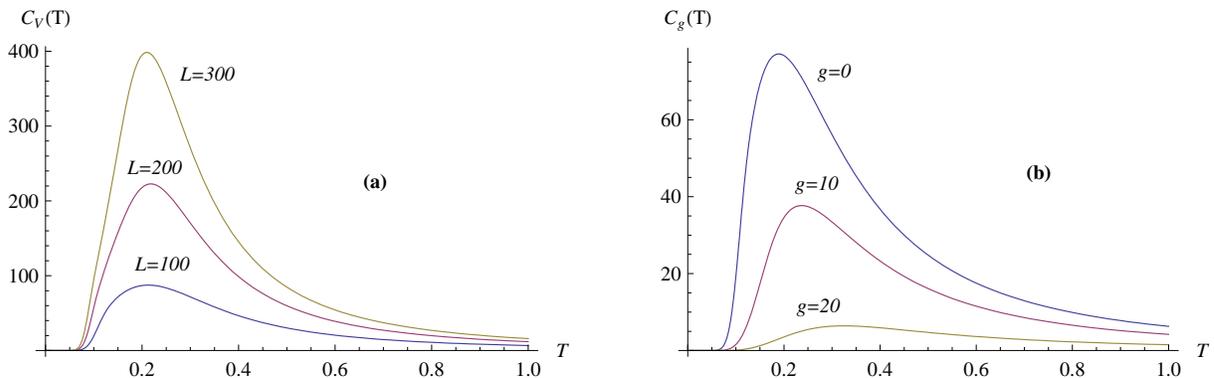}} \caption{ a) $C_V
(T)$ for $N=10$ and $L=100, 200, 300$. b) $C_g (T)$ for $L=100$
and $g=0,10,20$.}\label{figCv}
\end{figure}
\begin{figure}
\epsfysize=5.5cm \centerline{\epsfbox{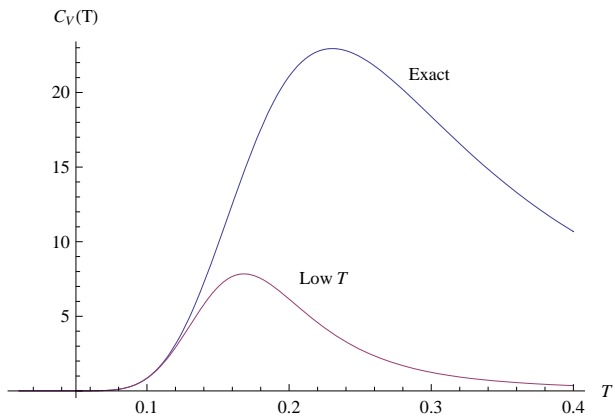}} \caption{
Exact and low-temperature approximation for the specific heat with
$L=20$ and $N=1$} \label{figCompCv}
\end{figure}

Next, we consider the specific heat at
constant volume (in this case the volume is the size of the chain
$V=L$):

 \beq
C_{V}(T)=-T \left(\frac{\partial^{2}F}{\partial T^{2}}\right)_{L}
\label{Cv}
 \eeq
\noindent
 The graph of $C_{V}$ against the temperature (Fig. \ref{figCv}
(a)) has the particular shape characteristic of the system with
finite energy levels (see comment after equation (\ref{Deg})). The
characteristic temperature $T_{ch}$ corresponds to the 
position of the peak in
$C_{V}$ showed in the graphs. For temperatures above and below
$T_{ch}$, the specific heat decreases rapidly. This well-known
behavior of the specific heat with temperature is known as the
Schottky anomaly \cite{Mandl,Pathria} and it is a general property 
of systems with energy levels with discrete degeneracy (see
$d_k$ above).
In the low temperature region, we have
 \beq
C_{V}(T<<1) \simeq k \left(\frac{\epsilon }{2k T}\right)^2
{\rm{sech}}^2\left(\frac{\epsilon }{2k T}-\ln
\sqrt{\frac{L(L-1)}{2}}\right) . \label{CvAprox}
 \eeq
\noindent
In this limit, the specific heat coincides with that for a
two-level system, since for low enough temperatures, the system
will only be able to access the ground state and the first excited
state. In figure \ref{figCompCv} , we plot the exact specific
heat from (\ref{Cv}) and the low-temperature approximation from
(\ref{CvAprox}).
Furthermore, we may define the topological specific heat as:
 \beq
C_{g}(T)=-T \left(\frac{\partial^{2}F_g}{\partial
T^{2}}\right)_{L, g} \label{Cvtop}
\eeq
where $F_g(L,T)=- \kappa T \ln(z_g (L,T))$. Note that $C_g$
can be identified with
the specific heat restricted to the diagrams on the topological
surface of genus $g$. In Fig. \ref{figCv} (b) we show that the
higher peaks correspond to the lower genera. This agrees with the
intuitive argument that considers a molecule with higher genus as 
strongly folded and, therefore, with reduced
number of degrees of freedom. We can carry out
the same analysis for $C_{V}$, given that the addition of a new
base molecule increases the number of degrees of freedom of the system.
In this sense, we can consider the relation $L \sim 1/g$.

In conclusion,
we have studied several thermodynamical and topological aspects
of the simplified model of RNA  of \cite{Zee1}.
We have presented an exact expression for the partition function
of the system, and gave an interpretation of the degeneracy of each
energy level as a function of $N$. Furthermore, we have calculated the
topological expansion of the partition function of the model, in
which the coefficients of the expansion can be interpreted as
the reduced partition functions for systems restricted to
topological surfaces of genus $g$. We showed that the
maximum genus of the configurations is $[L/4]$, for a molecule
of size $L$.
Moreover, we have calculated asymptotic expressions for some
thermodynamical observables, as a function of the temperature.
Analyzing the expressions for the chemical potential and
entropy, within our data, we find a consistent interpretation relating the variable
$1/N$ (arising from the matrix model) and the concentration of
$\rm{Mg}^{++}$, as it has
been suggested in \cite{Zee2,Zee5}.

MdE thanks to Mat\'{\i}as Reynoso for helpful discussions.

%
\def\NP{{\it Nucl. Phys.\ }}
\def\PRL{{\it Phys. Rev. Lett.\ }}
\def\PL{{\it Phys. Lett.\ }}
\def\PR{{\it Phys. Rev.\ }}
\def\CMP{{\it Comm. Math. Phys.\ }}
\def\IJMP{{\it Int. J. Mod. Phys.\ }}
\def\MPL{{\it Mod. Phys. Lett.\ }}
\def\RMP{{\it Rev. Mod. Phys.\ }}
\def\AP{{\it Ann. Phys.\ }}
\def\PRE{{\it Phys. Rev. E\ }}
\def\EPL{{\it Eur. Phys. Lett.\ }}

\end{document}